# Weak Decays of Heavy-Light Mesons on the Lattice: Semi-Leptonic Formfactors[*]

S. Güsken[a], G. Siegert[a,b], K. Schilling[a,b] [†]

[a]*Physics Department, Bergische Universität, Gesamthochschule Wuppertal*

*Gaußstrasse 20, D-42097 Wuppertal, Germany*

[b]*HLRZ c/o KFA Jülich, D-52425 Jülich, Germany*

*and DESY, Hamburg, Germany*

(August 4, 1995)

## Abstract

We report results (on an intermediate statistics sample) of a study of weak semi-leptonic formfactors of $B$ and $D$ decays, addressing the uncertainties from mass extrapolations to chiral and to heavy quarks. Moreover, we present a nonperturbative test to the LMK current renormalization scheme for vector current *transition* matrix elements and find remarkable agreement.

## I. INTRODUCTION

The determination of weak formfactors for the semi-leptonic decays of heavy-light mesons presents a big challenge for further exploration of the flavordynamics within the Standard Model, both experimentally *and* theoretically. The reason is that the study of weak interaction amplitudes like the Kobayashi-Maskawa parameters calls for good control of the hadronic binding effects in the analysis.

Heavy quark effective theory provides a suitable framework of understanding the 'geometric' part of the spin-flavor structure of heavy quark decay in the infinite mass limit

[†]Invited talk given at the German-Japanese workshop *Physics on Parallel Computers*, Yamagata 1995

[*]Work supported by DFG grant Schi 257/3-2 and 3-3

$M_h \to \infty$, but cannot predict the $1/M_h$ corrections in detail which are likely to be important in the mass regions of D- and B-mesons. For this reason great effort has been spent ever since the pioneering work of the El-Khadra et al [1] to compute QCD effects on these decays by means of lattice gauge theory.

In this contribution, we present a status report about our ongoing work on $24^3 \times 64$ lattices with quenched Wilson fermions at $\beta = 6.3$, based on 60 configurations (out of a total sample of 100 configurations) that we are carrying out on the 32 node connection machine CM5 with 40 Gbytes of parallel disc space SDA, at Wuppertal University. The light quarks were represented by the following set of hopping parameters $\kappa_l$: .1511, .1507, .1490, .1450, while the heavy quarks were varied over the set of $\kappa_h$: .1400, .1350, .1300, and .1200. The initial meson was chosen at rest, while the final meson could carry 11 different momenta (in lattice units): (0,0,0), (1,0,0), (1,1,0), (1,1,1), (2,0,0), and permutations.

## II. MASS EXTRAPOLATIONS

Let us for the moment consider the momentum transfer dependence of the formfactors, leaving aside the issue of the lattice renormalization of these quantities.

**Chiral extrapolation** Fig.1 gives an idea on the quality of our data, in terms of the conventional formfactors $f_+$ and $f_0$ for the transition $PS \to PS + leptons$ as well as the formfactors $A_1$, $A_2$, and $V$ for the decay $PS \to V + leptons$. For the formfactor definitions, see Ref. [3]. Kinematically, this figure refers to a heavy quark ($\kappa_i = .1350$ decaying into a light one ($\kappa = .1490$), with a spectator quark, $\kappa_{spec} = .1490$. The full lines represent the pole dominance predictions, with pole masses from the lattice two-point functions.

In Fig. 2 we demonstrate, for comparison the effects of chiral extrapolation for $\kappa_f$ and $\kappa_{spec}$. Physically this figure relates to the decays $D \to \rho$ and $D \to \pi$.

Unfortunately, our lattice allows only for a small variation in the momentum transfer between the initial and final state hadrons, $q^2$. Generally, we find *consistency* with (but no compelling evidence for) pole dominance.

**Extrapolation to $B$ Meson** It is well known that currently accessible lattice sizes do not allow for a direct simulation of $B$ decays and therefore require a substantial mass extrapolation, from the D to the B meson regime. While the transitions $B \to D, D^*$ lend themselves rather easily to a heavy quark effective theory (HQET) like analysis [7], the situation is much more involved in the decays $B \to \pi, \rho$. One option is to carry out this extrapolation in a manner suggested by HQET, i.e. at fixed value of $\omega = v_i v_f$ [3,8][1]. This extrapolation implies a big change in $q^2$, which has to be compensated by a substantial counter-extrapolation to $q^2 = 0$, the conventional point of comparison. Given the uncertainties in the formfactors this procedure introduces sizeable systematic errors. We therefore prefer to carry out the extrapolations at *fixed* $q^2 = 0$, using different functional forms of $M_{PS}$ dependence. This appears to be a better controlled way to judge the systematic errors on the final numbers.

In Fig.3 we display four of the formfactors $F(q^2 = 0)$ versus the inverse mass of the decaying meson, $M_{PS}^{-1}$. The final meson corresponds to $K$ and $K^*$, respectively. The differ-ent curves represent different types of linear extrapolations to the $B$ meson mass (leftmost points): full lines fits to $F$ itself, dashed lines fits to $F\sqrt{M_{PS}}$, dotted lines fits to $F\sqrt{M_{PS}}^{-1}$. The plots illustrate that the data is not sensitive yet to discriminate between the $M$ dependencies of the ansätze.

---

[1] In ref. [8] also an extrapolation in mass at $q^2 = 0$ has been performed.

### III. RENORMALIZATION

In order to relate our results to the continuum we need to renormalize the matrix elements of the local (non-conserved) vector and axial vector currents. It has been repeatedly pointed out that – on top of their short distance part – the renormalization factors may receive large $O(ma)$ contributions at large $ma$.

An obvious starting point is to take these contributions into account by approximately adjusting the normalization of the lattice quark propagator $\Delta(t, \vec{p} = \vec{0})$ to the one of the continuum quark propagator [2] which has been advocated and refined more recently by Lepage, Kronfeld and Mackenzie [5], and will hence be quoted as LMK scheme. From the free case relation $\Delta(t, \vec{p} = \vec{0})^{cont} = 2\kappa(1 + ma)\Delta(t, \vec{p} = \vec{0})^{latt}$ one uses mean field arguments to obtain

$$\Delta(t, \vec{p} = \vec{0})^{cont} \simeq 2\kappa N_{LMK}(\kappa)\Delta(t, \vec{p} = \vec{0})^{latt} \qquad (1)$$

for the interacting case. As pointed out in ref. [6]. the normalization factor $N_{LMK}(\kappa) = \frac{1}{2\kappa}(1 - \frac{3}{4}\frac{\kappa}{\kappa_c})$ can be "eaten up" by an additional time hop of the lattice quark propagator

$$\Delta(t, \vec{p} = \vec{0})^{cont} \simeq \Delta(t+1, \vec{p} = \vec{0})^{latt} . \qquad (2)$$

In this section we will measure the mass dependence of the renormalization constant of the time component of the vector current and compare the results to the LMK predictions as well as to the naive expectations, in which the LMK factor is absent.

In order to improve on statistical accuracy, we will work with ratios of 3-point functions, rather than those functions themselves. We emphasize that the mass dependence of such ratios directly measures lattice artefacts, as it disappears in the continuum limit.

We consider the ratio[2]

$$R_{H'H}(\vec{p}) = \frac{\langle PS(m_{H'}, m_l)|V_0^{cons}|PS(m_H, m_l)\rangle_{\vec{p}}}{\langle PS(m_{H'}, m_l)|V_0^{loc}|PS(m_H, m_l)\rangle_{\vec{p}}} . \qquad (3)$$

The ratio $R_{HH}(\vec{0})$ yields directly the renormalization factor of $V_0^{loc}$, as $V_0^{cons}$ is unaltered in this instance. For the study of decays we are rather interested in the ratios $R_{hH}(\vec{0})$ and $R_{HH}(\vec{p})$, however.

In order to derive predictions for the mass dependence of the ratios $R$ above we have to study the 3-point functions

$$G_{HH'l}^{cons(loc)}(0, y_0, x_0) = \sum_{\vec{x},\vec{y}} \langle P_{H'l}^\dagger(\vec{0}, 0)V_0^{cons(loc)}(\vec{y}, y_0)P_{Hl}(\vec{x}, x_0)\rangle , \qquad (4)$$

---

[2]A similar analysis has been carried out in ref. [9].

where $P$ denotes the operator of the pseudoscalar meson and

$$V_0^{loc}(\vec{y}, y_0) = \bar{\Psi}_{H'}(\vec{y}, y_0)\gamma_0 \Psi_H(\vec{y}, y_0) \tag{5}$$

$$V_0^{cons}(\vec{y}, y_0) = \frac{1}{2}\left[\bar{\Psi}_{H'}(\vec{y}, y_0)(\gamma_0 - 1)U_0(\vec{y}, y_0)\Psi_H(\vec{y}, y_0 + 1)\right.$$
$$\left. + \bar{\Psi}_{H'}(\vec{y}, y_0 + 1)(\gamma_0 + 1)U_0^\dagger(\vec{y}, y_0)\Psi_H(\vec{y}, y_0)\right] . \tag{6}$$

For $0 \ll y_0 \ll x_0$ one finds $R_{HH'} = G_{HH'l}^{cons}/G_{HH'l}^{loc}$. In terms of quark propagators, $\Delta$, one easily obtains

$$G_{HH'l}^{loc}(0, y_0, x_0) = \langle \sum_{\vec{x},\vec{y}} Tr\left\{\Delta_H(\vec{y}, y_0, \vec{x}, x_0)\gamma_5 \Delta_l(\vec{x}, x_0, \vec{0}, 0)\Delta_{H'}^\dagger(\vec{y}, y_0, \vec{0}, 0)\gamma_5\gamma_0\right\}\rangle \tag{7}$$

$$G_{HH'l}^{cons}(0, y_0, x_0) = \langle \frac{1}{2} \sum_{\vec{x},\vec{y}} [ \tag{8}$$
$$Tr\left\{\Delta_H(\vec{y}, y_0 + 1, \vec{x}, x_0)\gamma_5 \Delta_l(\vec{x}, x_0, \vec{0}, 0)\Delta_{H'}^\dagger(\vec{y}, y_0, \vec{0}, 0)\gamma_5(\gamma_0 - 1)U_0(\vec{y}, y_0)\right\}$$
$$+ Tr\left\{\Delta_H(\vec{y}, y_0, \vec{x}, x_0)\gamma_5 \Delta_l(\vec{x}, x_0, \vec{0}, 0)\Delta_{H'}^\dagger(\vec{y}, y_0 + 1, \vec{0}, 0)\gamma_5(\gamma_0 + 1)U_0^\dagger(\vec{y}, y_0)\right\}]\rangle .$$

Inserting the relations (1) and (2) for the quark propagators one can now read off the LMK predictions for the mass dependence of the ratios $R(\vec{0})$. Note that the second term in eq.(8) is expected to be negligible within our mass range[3]. Because of $y_0 - x_0 < 0$ one gets $\Delta_H(y_0, x_0) \sim (1 - \gamma_0)$ for very heavy quarks, and the second term becomes proportional to $(1 - \gamma_0)(1 + \gamma_0) = 0$. The LMK predictions for the ratios therefore read

$$R_{HH}(\vec{0}) = \tilde{Z}_V^{loc}\frac{1 - \frac{3}{4}\frac{\kappa_H}{\kappa_c}}{2\kappa_H} \qquad R_{hH}(\vec{0}) = \frac{\tilde{Z}_V^{loc}}{\tilde{Z}_V^{cons}}\frac{1 - \frac{3}{4}\frac{\kappa_H}{\kappa_c}}{2\kappa_H} , \tag{9}$$

where $\tilde{Z}_V^{loc}$ and $\tilde{Z}_V^{cons}$ are constants. We emphasize that $R_{hH}(\vec{0})$ only depends on $m_H$ but not on $m_h$. This is a consequence of the cancellation of the second term in eq.(8). The naive prediction simply reads $R_{HH}(\vec{0}) = Z_V = R_{hH}$.

In fig.4 we compare our numerical results for $R_{HH}(\vec{0})$ with the LMK and the naive predictions. We have inserted the (1-loop) perturbative values $\tilde{Z}_V^{loc} = 1 - 0.82/4\pi g^2$, $Z_V^{loc} = 1 - 0.174081g^2$, using the boosted coupling and assumed $\tilde{Z}_V^{cons} = Z_V^{cons} = 1$. Throughout our mass range of $m_H$ we find the LMK prediction to approximate our data within 5%. The naive prediction, however, fails drastically.

For actual physics applications it is more interesting to study the situation, where initial and final nonspectator quarks carry different masses. For that matter, fig.4 shows the ratio $R_{hH}(\vec{0})$ as a function of $m_h$ at fixed $m_H$. Again we find good agreement with the LMK prediction. This finding has important bearing on the reliability of our calculations with *light* final states.

---

[3]This has also been verified numerically from our data.

There remains the final issue, whether LMK holds as well in the case of unequal momenta of initial and final state hadrons. To this end, fig.4 displays the ratio $R_{HH}(\vec{p})$ on the momentum set $|\vec{p}a| = 2\pi/64 \times \{0, 1, \sqrt{2}, 2\}$. The solid line represents the above LMK prediction, for $\vec{p} = \vec{0}$. We find that the data moves only marginally with increasing $\vec{p}$ and is therefore well approximated by the LMK prediction over the entire $\vec{p}$ range. This result is not at all trivial, as one would expect the normalization factor to be affected by the momentum.

## IV. PHYSICS RESULTS

In table I we have collected our results for the B decays into charmed particles, while the tables III and II contain our results on decays into light hadrons and on D decays, respectively, together with the data of other lattice investigations, as well as model and sum rule calculations.

We find good agreement with experimental data for the $D \to K, K^*$ decays, with possible exception of $A_2(0)$, where the errors are large. Note that the kinematical constraint $f_+(0) = f_0(0)$ is well fulfilled by our data!

While our results for $D \to \pi, \rho$ are generally fairly consistent with the results from other lattice investigations, there is still rather large scatter in the results for the decays $B \to \pi, \rho$. This is related to the heavy mass extrapolations. We have not imposed the constraint $f_+(0) = f_0(0)$ in our analysis. Therefore the difference in these quantities reflects systematic uncertainties. The extrapolative situation for these formfactors can be traced from the upper plots in fig.3. $f_0$ is determined with higher precision, due to the kinematical point $\vec{p}_f = \vec{0}$. A fit using the constraint $f_+(0) = f_0(0)$ along the mass trajectory would favour the value 0.20, of course. A similar kind of uncertainty is borne out by the range quoted recently by the UKQCD collaboration [8] for $f_+(0) = f_0(0)$. They have used the latter constraint within an HQET inspired extrapolation procedure, using several variants for the $q^2$-dependence of the formfactors.

At this stage, one should note, that reliable knowledge obout the Z-factors is urgently needed, as they enter directly into the lattice predictions for the formfactors. It is therefore of great interest to continue studies about nonperturbative lattice renormalization.

FIGURES

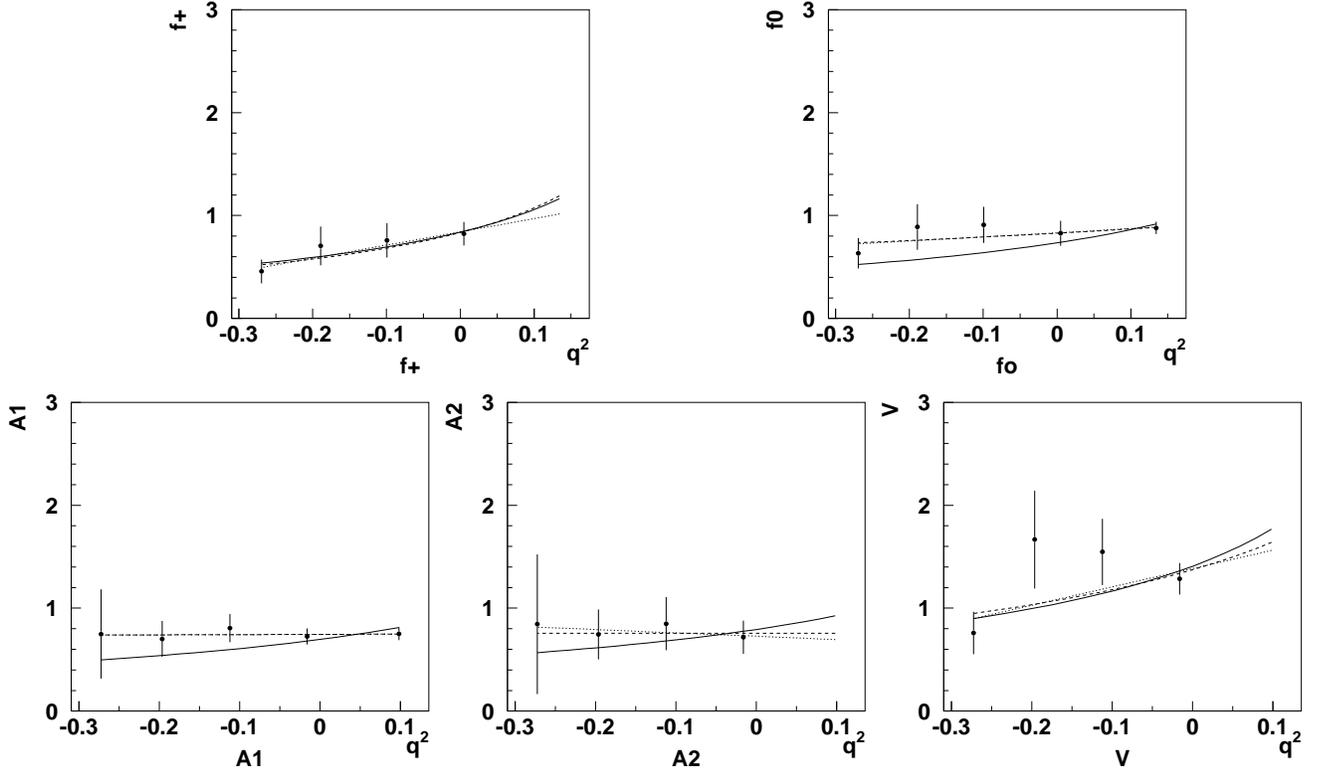

FIG. 1. $q^2$ dependence of the formfactors, at $\kappa_i = .1350$, and $\kappa_f = \kappa_{spec} = .1490$. The curves refer to fits: pole dominance with mass from two-point function (full line), with mass fitted to formfactor (dashed line), and linear fit (dotted line).

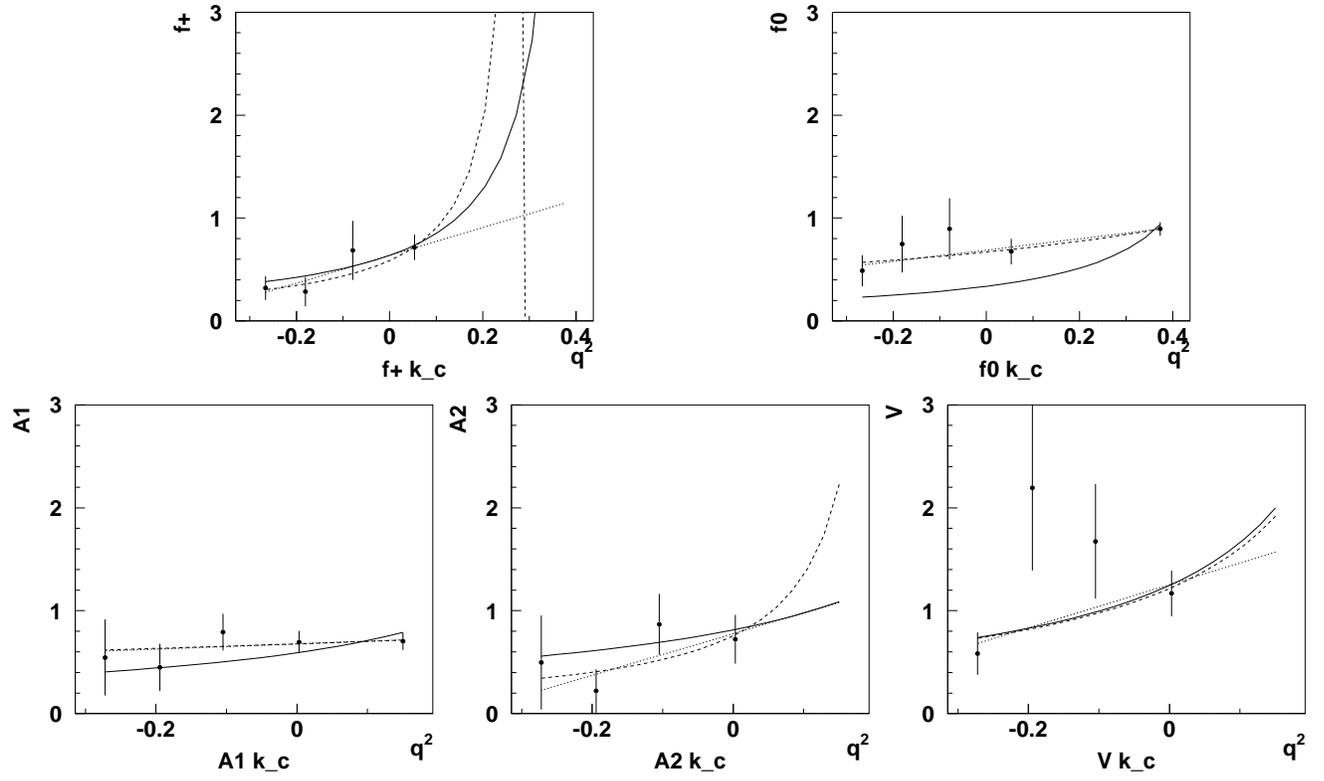

FIG. 2. Same as Fig.1, but extrapolated to the chiral limit.

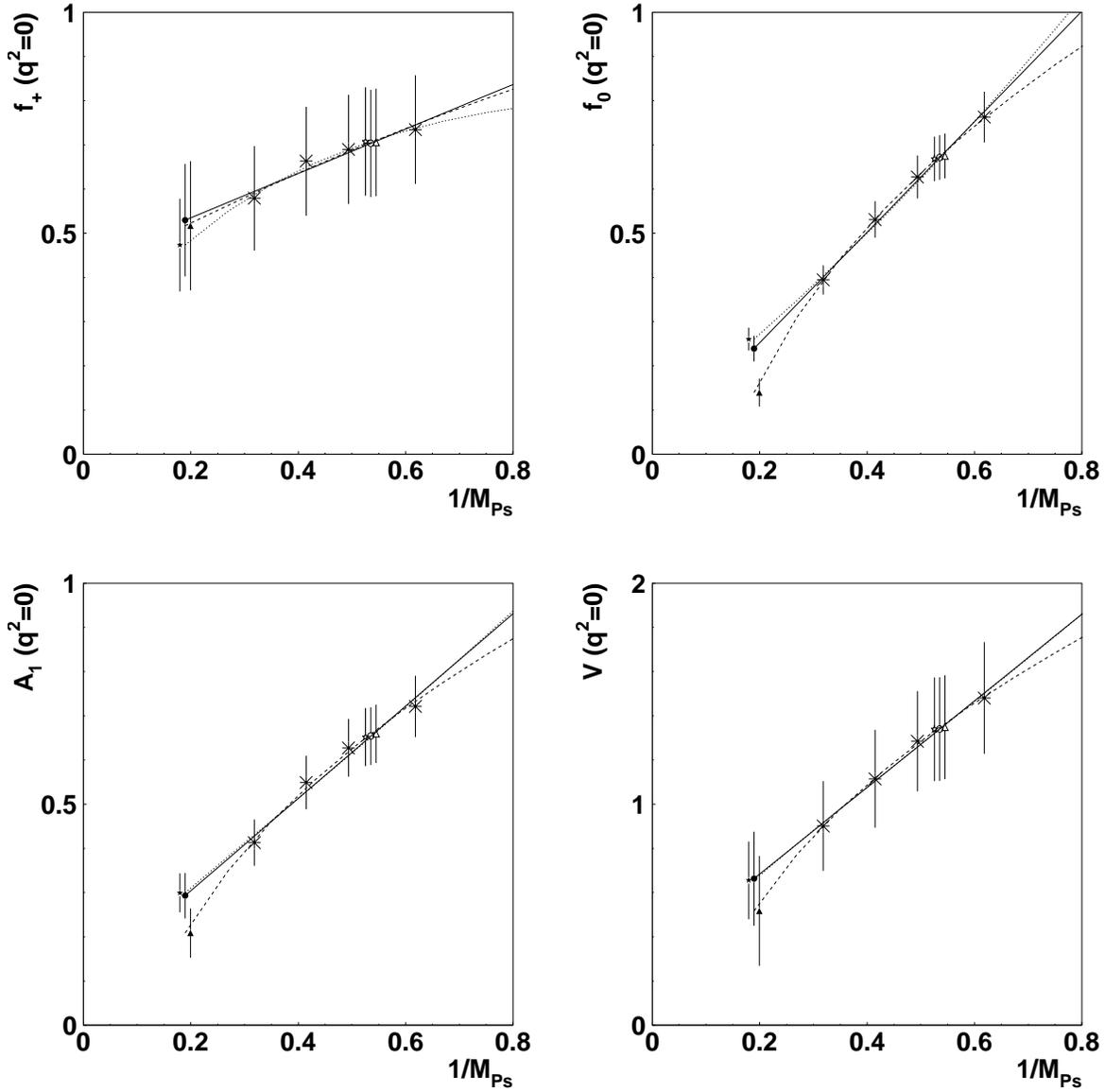

FIG. 3. Extrapolation of formfactors in the mass of the decaying pseudoscalar particle [in $GeV$], with final state hadron $K$ and $K*$, respectively. The data are included as crosses, and the extrapolated points are plotted at the inverse B-mass, $M_{PS}^{-1} = .2$. All fits are linear: (a) to F itself (full line), (b) to $FM_{PS}^{1/2}$ (dashed line), (c) to $FM_{PS}^{-1/2}$ (dotted lines). The interpolated results at the D-mass are included for convenience (open symbols).

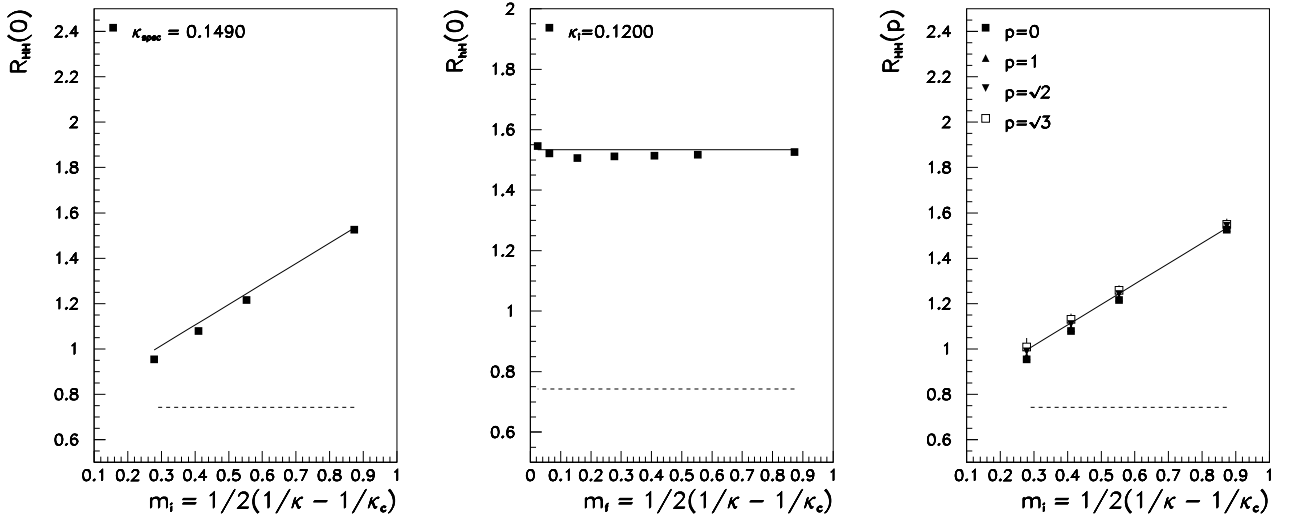

FIG. 4. The ratios $R_{HH}(\vec{0})$, $R_{hH}(\vec{0})$ and $R_{HH}(\vec{p})$ as a function of the quarkmass. Solid lines represent the LMK and dashed lines the *naive* prediction.

TABLES

TABLE I. Formfactors for $B \to D$, $D^*$ und $B_s \to D_s$, $D_s^*$.

| $B \to D$, $D^*$ | | | | |
|---|---|---|---|---|
| $f_+(0)$ | $f_0(0)$ | $V(0)$ | $A_1(0)$ | $A_2(0)$ |
| 1.01(60) | 0.88(29) | 1.84(1.10) | 0.97(32) | 2.76(1.58) |

| $B_s \to D_s$, $D_s^*$ | | | | |
|---|---|---|---|---|
| $f_+(0)$ | $f_0(0)$ | $V(0)$ | $A_1(0)$ | $A_2(0)$ |
| 0.99(42) | 0.85(24) | 1.84(1.10) | 0.91(24) | 2.25(1.12) |

TABLE II. Formfactors at $q^2 = 0$ for the decays $D \to K$ und $D \to K^*$.

|  | Method | $f_+(0)$ | $f_0(0)$ | $V(0)$ | $A_1(0)$ | $A_2(0)$ |
|---|---|---|---|---|---|---|
| World av. [10] | Exp't | 0.77(4) |  | 1.16(16) | 0.61(5) | 0.45(9) |
| This work |  | $0.71(12)^{+10}_{-7}$ | $0.67(6)^{+9}_{-7}$ | $1.34(24)^{+19}_{-14}$ | $0.61(6)^{+9}_{-7}$ | $0.83(20)^{+12}_{-8}$ |
| LMMS [11] | Latt. | 0.63(8) |  | 0.86(10) | 0.53(3) | 0.19(21) |
| BKS [12] | Wilson | $0.90^{+8+21}_{-8-21}$ | $0.70^{+8+24}_{-8-24}$ | $1.43^{+45+48}_{-45-49}$ | $0.83^{+14+28}_{-14-28}$ | $0.59^{+14+24}_{-14-23}$ |
| ELC [3] |  | 0.60(15)(7) |  | 0.86(24) | 0.64(16) | 0.40(28)(4) |
| BG [13] |  | 0.63(3) | 0.66(2) | 1.18(7) | 0.65(3) | 0.41(7) |
| APE [14] | Latt. | 0.72(9) |  | 1.0(2) | 0.64(11) | 0.46(34) |
| UKQCD [15] | Clover | $0.67^{+7}_{-8}$ | 0.65(7) | $1.01^{+30}_{-13}$ | $0.70^{+7}_{-10}$ | $0.66^{+10}_{-15}$ |
| ISGW [16] |  | 0.76 − 0.82 |  | 1.1 | 0.8 | 0.8 |
| WSB [17] | Quark- | 0.76 |  | 1.27 | 0.88 | 1.15 |
| KS [18] | Models | 0.76 |  | 0.8 | 0.82 | 0.8 |
| GS [19] |  | 0.69 |  | 1.5 | 0.73 | 0.55 |
| BBD [20] |  | $0.60^{+15}_{-10}$ |  | 1.10(25) | 0.50(15) | 0.60(15) |
| AEK [21] | Sumrules | 0.60(15) |  |  |  |  |
| DP [21] |  | 0.75(5) |  |  |  |  |

TABLE III. Formfactors at $q^2 = 0$ for the decays $D$ and $B$ to $\pi$ and $\rho$.

$D \to \pi$ und $D \to \rho$

|  | Method | $f_+(0)$ | $f_0(0)$ | $V(0)$ | $A_1(0)$ | $A_2(0)$ |
|---|---|---|---|---|---|---|
| This work |  | $0.68(13)^{+10}_{-7}$ | $0.65(5)^{+9}_{-7}$ | $1.31(25)^{+18}_{-13}$ | $0.59(7)^{+8}_{-6}$ | $0.83(20)^{+12}_{-8}$ |
| LMMS [11] | Latt. | $0.58(9)$ |  | $0.77(9)$ | $0.47(7)$ | $-0.07(42)$ |
| BKS [12] | Wilson | $0.84(12)(35)$ | $0.70(8)(24)$ | $1.07(49)(35)$ | $0.65(15)^{+24}_{-23}$ | $0.59^{+28}_{-25}$ |
| BG [13] |  | $0.52(6)$ | $0.57(4)$ | $1.12(11)$ | $0.58(4)$ | $0.34(10)$ |
| UKQCD [15] | Clover | $0.61^{+12}_{-11}$ |  | $0.95^{+29}_{-14}$ | $0.63^{+6}_{-9}$ | $0.51^{+10}_{-15}$ |
| ISGW [16] |  | $0.51$ | $0.51$ |  |  |  |
| WSB [17] | Quark- | $0.69$ |  | $1.23$ | $0.78$ | $0.92$ |
| KS [18] | Models | $0.69$ |  | $1.23$ | $0.78$ | $0.92$ |
| AEK, DP [21] | Sumrules | $0.6$–$0.75$ | $0.6$–$0.75$ |  |  |  |

$B \to \pi$ und $B \to \rho$

|  | Method | $f_+(0)$ | $f_0(0)$ | $V(0)$ | $A_1(0)$ | $A_2(0)$ |
|---|---|---|---|---|---|---|
| This work | Latt. | $0.50(14)^{+7}_{-5}$ | $0.20(3)^{+2}_{-3}$ | $0.61(23)^{+9}_{-6}$ | $0.16(4)^{+22}_{-16}$ | $0.72(35)^{+10}_{-7}$ |
| ELC a [3] | Wilson | $0.28(14)$ |  | $0.37(14)$ | $0.24(6)$ | $0.39(24)$ |
| ELC b [3] |  | $0.33(17)$ |  | $0.40(16)$ | $0.21(5)$ | $0.47(28)$ |
| APE a [14] | Latt. | $0.29(6)$ |  | $0.45(22)$ | $0.29(16)$ | $0.24(56)$ |
| APE b [14] | Clover | $0.35(8)$ |  | $0.53(31)$ | $0.24(12)$ | $0.27(80)$ |
| UKQCD [8] |  | $0.24^{+4}_{-3}$ | $0.24^{+4}_{-3}$ |  | $0.27^{+7}_{-4} \pm 3$ |  |
| ISGW [16] | Quark- | $0.09$ |  | $0.27$ | $0.05$ | $0.02$ |
| WSB [17] | Models | $0.33$ |  | $0.33$ | $0.28$ | $0.28$ |
| Ball [22] | Sumrules | $0.26(2)$ |  | $0.6(2)$ | $0.5(1)$ | $0.4(2)$ |